\documentclass[12pt,onecolumn]{IEEEtran}
\usepackage{cite}
\usepackage[final]{graphicx}
\usepackage{amsmath,amsfonts}
\usepackage{caption}

\newcommand*\mystrut[1]{\vrule width0pt height0pt depth#1\relax}
\newtheorem{lemma}{Lemma}
\newtheorem{theorem}{Theorem}

\newtheorem{remark}{Remark}
\newtheorem{corollary}{Corollary}

\def\figref#1{Figure~\ref{#1}}
\def\apref#1{Appendix~\ref{#1}}
\def\secref#1{Section~\ref{#1}}

\def\coref#1{Corollary~\ref{#1}}

\def\QED{{\ensuremath \Box}}

\def\bx{\mathbf x}
\def\bf{\mathbf f}
\def\bu{\mathbf u}
\def\bq{\mathbf q}
\def\bp{\mathbf p}
\def\by{\mathbf y}
\def\bs{\mathbf s}
\def\bz{\mathbf z}

\def\bh{\mathbf h}

\def\bA{\mathbf A}
\def\bW{\mathbf W}
\def\bT{\mathbf T}
\def\bB{\mathbf B}
\def\bU{\mathbf U}
\def\bI{\mathbf I}

\def\bnu{\boldsymbol \nu}

\def\bzero{\mathbf 0}
\def\hk{{\hat k}}
\def\cD{{\mathcal D}}
\def\cDin{{\mathcal D}^{(\text{in})}}

\def\cC{{\mathcal C}}
\def\cK{{\mathcal K}}
\def\cJ{{\mathcal J}}
\def\cN{{\mathcal N}}
\def\cM{{\mathcal M}}
\def\cW{{\mathcal W}}
\def\cT{{\mathcal T}}
\def\hj{{\hat j}}
\def\hk{{\hat k}}
\def\hr{{\hat r}}

\def\hht{{\hat t}}
\def\setN{{\mathbb N}}
\def\setZ{{\mathbb Z}}
\def\setR{{\mathbb R}}
\def\bH{\mathbf H}
\def\size #1x#2{$#1\times #2$}
\newcommand*{\bydef}{\mathrel{\vcenter{\baselineskip0.5ex \lineskiplimit0pt
   \hbox{\scriptsize.}\hbox{\scriptsize.}}}=}
 
\def\paren#1{\left(#1\right)}
\def\part#1{\medskip\noindent\textsc{#1}}
\def\st{\,\bigr|\,}
\def\thref#1{Theorem~\ref{#1}}
\def\leref#1{Lemma~\ref{#1}}
\def\reref#1{Remark~\ref{#1}}
\def\recol#1{Corollary~\ref{#1}}

\normalsize

\begin{document}

\title{Multiple-Antenna Interference Channels with
Real Interference Alignment and Receive Antenna Joint Processing}

\author{
  \IEEEauthorblockN{Mahdi Zamanighomi and Zhengdao Wang}\\
  \IEEEauthorblockA{Department of Electrical and Computer Engineering\\
  Iowa State University, Ames, Iowa, USA\\
  Email: \{mzamani,zhengdao\}@iastate.edu}
}

\maketitle

\begin{abstract}
In this paper, the degrees of freedom (DoF) regions of constant coefficient
multiple antenna interference channels are investigated. First, we consider a
$K$-user Gaussian interference channel with $M_k$ antennas at transmitter $k$,
$1\le k\le K$, and $N_j$ antennas at receiver $j$, $1\le j\le K$, denoted as a
$(K,[M_k],[N_j])$ channel. Relying on a result of simultaneous Diophantine
approximation, a real interference alignment scheme with joint receive antenna
processing is developed. The scheme is used to obtain an achievable DoF
region. The proposed DoF region includes two previously known results as
special cases, namely 1) the total DoF of a $K$-user interference channel with
$N$ antennas at each node, $(K, [N], [N])$ channel, is $NK/2$; and 2) the
total DoF of a $(K, [M], [N])$ channel is at least $KMN/(M+N)$. We next
explore constant-coefficient interference networks with $K$ transmitters and
$J$ receivers, all having $N$ antennas. Each transmitter emits an independent
message and each receiver requests an arbitrary subset of the messages.
Employing the novel joint receive antenna processing, the DoF region for this
set-up is obtained. We finally consider wireless X networks where each node is
allowed to have an arbitrary number of antennas. It is shown that the joint
receive antenna processing can be used to establish an achievable DoF region,
which is larger than what is possible with antenna splitting. As a special
case of the derived achievable DoF region for constant coefficient X network,
the total DoF of wireless X networks with the same number of antennas at all
nodes and with joint antenna processing is tight while the best inner bound
based on antenna splitting cannot meet the outer bound. Finally, we obtain a
DoF region outer bound based on the technique of transmitter grouping.


\noindent \rule[-8pt]{.5\textwidth}{.5pt}\\ The work has been presented in
part at the IEEE ISIT 2013 Conference.
\end{abstract}

\clearpage

\linespread{1.66}\normalsize
\section{Introduction} \label{intro} Characterizing the capacity region of
interference networks is a fundamental problem in information theory. Despite
remarkable progress in recent years, the capacity region of interference
networks remains unknown in general. Recent work has proposed to use degrees
of freedom (DoF) to approximate the capacity region of interference networks.
The DoF of a message is its rate normalized by the capacity of single-user
additive white Gaussian noise channel, as the signal-to-noise ratio (SNR)
tends to infinity. The DoF region quantifies the shape of the capacity region
at high SNR; see e.g., \cite{mamk06c,caja08}.

DoF investigations have motivated several fundamental ideas such as
interference alignment. With interference alignment, the interference signals
at any receiver from multiple transmitters are aligned in the signal space, so
that the dimensionality of the interference in the signal space can be
minimized. The remaining space is interference free and can be used for the
desired signals. Two commonly used alignment schemes are vector alignment and
real alignment \cite{jafa11,mgmk09}. In vector alignment, any transmit signal
is a linear combination of some vectors in a manner that the coefficients of
the linear combination carry useful data. This scheme designs the vectors so
that the interferences at each receiver are packed into a common subspace. The
orthogonal complement can be used for detecting useful data symbols. In real
alignment, the concept of linear independence over the rational numbers
replaces the more familiar vector linear independence. A Groshev type theorem
is usually used to guarantee the required decoding performance.

\subsection{DoF of interference channel}

DoF characterizations have been investigated for a variety of wireless
networks such as $K$-user interference channel and wireless X network. In the
$K$-user interference channel, the $k$-th transmitter has a message intended
for the $k$-th receiver. At receiver $k$, the messages from transmitters other
than the $k$-th are interference. The DoF region of the $K$-user interference
when all nodes are provided with the same number of antennas is known
\cite[Corollary~2]{krwy12i}.

In \cite{goja10}, Gou and Jafar studied the total DoF of the \size{M}x{N}
$K$-user interference channel where each transmitter has $M$ antennas and each
receiver has $N$ antennas. They showed the exact total DoF value is
$K\frac{MN}{M+N}$ under the assumption that
$R\bydef\frac{\max{(M,N)}}{\min{(M,N)}}$ is an integer and $K\geq R$. In
\cite{ghmk10c}, Ghasemi et al. employ antenna splitting argument to derive the
total DoF $K\frac{MN}{M+N}$ for fixed channels, which is optimal if
$K\geq\frac{M+N}{\text{gcd}(M,N)}$ even when $R$ is not an integer. In such
antenna splitting arguments, no cooperation is used either at the transmitter
side or at the receiver side. The outer bounds of these cases are based on
cooperation among groups of transmitters and receivers and employing the DoF
outer bound for 2-user multiple-input multiple-output (MIMO) interference
channel obtained in \cite{jafa07}. Note that the outer bound discussion is
regardless of whether the channel coefficients are constant or time-varying.

A novel genie chains approach for the DoF outer bound of \size{M}x{N} $K$-user
interference channel has been recently presented in \cite{wasj12}. In this
approach, a chain of mapping from genie signals provided at a receiver to the
exposed signal spaces of the receiver is served as the genie signals for the
next receiver until a genie with an acceptable number of dimensions is
obtained. As a result, it is proved that for any $K\geq 4$, the total DoF is
outer bounded by $K\frac{MN}{M+N}$ as long as $R\geq\frac{K-2}{K^2-3K+1}$.

The DoF region of MIMO $K$-user interference channels has not been obtained in
general for arbitrary number of antennas except for the $2$-user case
\cite{jafa07}.

\subsection{DoF of X network}

There is also increasing interest in characterizing DoF region of MIMO X
networks. A \size{K}x{J} MIMO X network consists of $K$ transmitters and $J$
receivers where each transmitter has an independent message for each receiver.
Notably, the X networks include interference channels as a special case.

The best known inner bounds on the total DoF of \size{K}x{J} MIMO time-varying
X networks with $N$ antennas at each node are based on:
\begin{enumerate}
\item Antenna splitting with no cooperation \cite{caja09}: The achievable
total DoF is attained by decomposing all transmitter and receiver antennas in
which we have an \size{NK}x{NJ} user single-input single-output X network.
Therefore, the best total DoF $N\frac{KJ}{K+J-\frac{1}{N}}$ is achieved.
However, there is a gap between the inner bound and the DoF outer bound,
$N\frac{KJ}{K+J-1}$, implying that a cooperation structure might be needed
here.
\item Joint signal processing \cite{sggj12}: Doing joint processing at either
transmitter or receiver side, the desired signals at any receiver can be
efficiently resolved from the interference. This new insight closes the
mentioned gap and the total DoF value $N\frac{KJ}{K+J-1}$ is achieved.
\end{enumerate}
These results offer an opportunity to revise our understanding of antenna
splitting technique. Independent processing at each antenna was initially
employed to simplify the achievability scheme of $K$-user MIMO interference
channels, which turned out to be optimal in some cases. However, as observed
in \cite{sggj12} allowing cooperation among antennas is essential for
establishing the desired DoF.

In the class of real interference alignment, the DoF of time-invariant
\size{K}x{J} MIMO X networks has not been studied to the best of our
knowledge. Also, except for the two-user case \cite{jash08}, the DoF region of
MIMO X networks when each node has an arbitrary number of antennas has not
been considered yet.

\subsection{Summary of Results}

In this paper, we employ recent results from the field of simultaneous
Diophantine approximation for systems of $m$ linear forms in $n$ variables to
analyze the performance of joint receive antenna processing. Based on the
analysis, we characterize the DoF region of several classes of time-invariant
multiple antenna interference networks.

To introduce the main concepts, we first study a time-invariant $K$-user MIMO
Gaussian interference channel with $N$ antennas at each node. We develop a
novel real interference alignment scheme for this channel and establish the
total DoF for this channel (\thref{th.KNN}).

Next, we focus on $K$-user MIMO Gaussian interference equipped with $M$
antennas at each transmitter and $N$ antennas at each receiver. For this
scenario, an achievable DoF region is established (\thref{th.KMN}). It is
shown that the achieved DoF region includes the previously known results as
special cases. We also establish an achievable DoF region for the $K$-user
MIMO Gaussian interference such that each node has an arbitrary number of
antennas (\thref{th.region.KMkNj}).

We then consider \size{K}x{J} MIMO interference network with general message
demands under assumption that all nodes have the same number of antennas. In
this model, each transmitter conveys an independent message and each receiver
requests an arbitrary subset of messages. With joint receive antenna
processing and real interference alignment, the exact DoF region is
established (\thref{th.region.KJNNW}).

We also apply our new scheme to the \size{K}x{J} MIMO X network and derive an
achievable DoF region (\thref{th.region.KJMkNj}), which is shown to be tight
under certain circumstances.

Finally, we discuss the outer bound in \secref{sec.outer}. By suitable
transmitter grouping argument, we obtain an outer bound on the DoF region for
a $K$ user interference channel with $M$ antennas at every transmitter and $N$
antennas at every receiver (\thref{th.outer}).

Notation: Throughout the paper, $K$, $J$, $M$, $N$, $D$, and $D'$ are integers
and $\cK=\{1, \ldots, K\}$, $\cJ=\{1, \ldots, J\}$, $\cM=\{1, \ldots, M\}$,
$\cN=\{1, \ldots, N\}$. We use $k$, $\hk$ as transmitter indices, and $j$,
$\hj$ as receiver indices. Superscripts $t$ and $r$ are used for transmitter
and receiver antenna indices. Letters $i$ and $l$ are used as the indices of
directions and streams (to be specified later), respectively. The set of
integers, positive integers, and real numbers are denoted as $\setZ$, $\setN$,
and $\setR$, respectively. The set of non-negative real numbers is denoted as
$\setR_+$. For a positive integer $Q$, we define $\setZ_Q\bydef\{z|z\in \setZ,
-Q\le z\le Q\}$. We denote the set of directions, a specific direction, and
the vector of directions using $\cT$, $T$, and $\bT$ respectively. Vectors and
matrices are indicated by bold symbols. We use $[M_k]_{k=1}^K$ to denote
vector $(M_1,\ldots,M_K)$, and $[d_{j,k}]_{j=1,k=1}^{J,K}$ the \size{J}x{K}
matrix with element $d_{j,k}$ in the $(j,k)$th position. When there is no
confusion, $[M_k]$ is used as an abbreviation for $[M_k]_{k=1}^K$, and $[M]$
is used to denote a vector where all $M_k$ are equal to $M$. We use
$(\cdot)^*$ to denote matrix transpose, $\otimes$ the Kronecker product of two
matrices, $\cup$ union of sets, ${\|\bx\|}_{\infty}$ the infinity norm of
vector $\bx$, and ${\|\bx\|}_{2}$ the 2-norm of vector $\bx$.

\section{Diophantine approximation and joint receive antenna processing}
\label{sec.dio}

The problem of Diophantine approximation is to approximate real numbers with
rational numbers. Let $a/b$ denote a rational approximation to a real number
$\omega$. It is useful to identify upper and lower bounds of $\lvert
\omega-a/b\rvert$, as a function of $b$. In addition to approximating a single
real number, simultaneous approximations to several rational numbers can be
considered. The problem of simultaneous Diophantine approximation is to
identify for a given real $n\times m$ matrix $\bA$, how small the distance
from $\bA \bq$ to $\setZ^n$, in terms of $\bq\in \setZ^m$, can be made
\cite{klmw10}.

To see how simultaneous Diophantine approximation can be useful in
communications, consider a communication receiver that receives a vector of
signals, $\by$, in the following form:
\begin{equation}\label{eq.basic}
  \by = \bA \bx + \bnu
\end{equation}
where $\bA$ is a real $n\times m$ matrix, and $\bx\in\setR^{m\times 1}$
contains information symbols to detected, and $\bnu\in \setR^{n\times 1}$ is
additive noise, assumed to contain independent and identically distributed
zero-mean Gaussian random variables. If we choose $\bx\in \{\lambda
\setZ_Q^m\}$ where $\setZ_Q^m = \{(q_1,\ldots,q_m)|q_i\in\setZ_Q, 1\leq i \leq
m\}$ and $\lambda$ is a positive real number that can be used to control the
signal power, then the block error probability for detecting $\bx$ is
determined by the set of distances $\{\| \bA(\bx-\bx')\|_2
\,\bigr|\,\bx,\bx'\in \lambda \setZ_Q^m\}$. Therefore, an upper bound on this
error probability can be obtained by lower bounding $\|\bA\bq\|_2$, over
non-zero $\bq\in \setZ^m$.

In this paper, the dimensionality $n$ of $\bA$ will be the number of receive
antennas. However, the other dimension $m$ is in general much larger than the
total number of transmit antennas. The signal $\bx$ will contain useful
information from the intended transmitters, as well as the interference
signals from unintended transmitters. Our strategy will be to select suitably
scaled integer lattice constellation for $\bx$, create the equivalent matrix
$\bA$ through transmitter designs that align the interferences at the
receivers, and perform joint processing of the entries of $\by$ for detecting
$\bx$. The fact that signals in $\by$ are jointly processed embodies what we
term as joint receive antenna processing.

It is known that for almost every $\bA$ in the Lebesgue sense, for any
$\delta>0$, there are at most finitely many $\bq\in \setZ^m$ with (see e.g.,
\cite[Sec.~1]{klmw10})
\begin{equation}
\|\bA \bq - \bp\|_\infty<\|\bq\|_\infty^{-m/n-\delta} \text{ for some } \bp\in
\setZ^n.
\end{equation}
Therefore, for almost every $\bA$, there are at most finite $\bq$ such that
$\|\bA\bq\|_\infty <\|\bq\|_\infty^{-m/n-\delta}$. If we further restrict
$\bA$ to be such that elements on at least one row are \emph{rationally
independent}, meaning no element can be written as a linear combination of the
other elements with rational coefficients, then for large enough $Q$,
$\|\bA\bq\|>Q^{-m/n-\delta}$ for all non-zero $\bq\in \setZ_Q^m$. Note that
imposing the rational independence requirement only removes a set of $\bA$ of
zero Lebesgue measure.

In our communication system design, the elements of $\bA$ are functionally
dependent. We will rely on the result of \cite[Theorem~1.2]{klmw10}, which we
state below as a lemma in a slightly different form that is suitable for its
application to communication problems. See \apref{ap.nondegen} regarding
non-degeneracy of manifolds. The proof of the lemma is provided in
\apref{ap.manifold}.

\begin{lemma}\label{le.manifold}
Let $\bf_i$, $i=1,\ldots,n$ be a non-degenerate map from an open set $U_i
\subset\setR^{d_i}$ to $\setR^m$ and
\begin{equation} \nonumber
\bA : U_1 \times \ldots \times U_n \rightarrow \cM_{n,m}, \quad
(\bh_1,\ldots,\bh_n)\longmapsto \left( \begin{array}{ccc}
\bf_1(\bh_1)  \\
\vdots \\
\bf_n(\bh_n)  \end{array} \right)
\end{equation}
where $\cM_{n,m}$ denotes the space of $n\times m$ real matrices. Then, for
almost all $(\bh_1, \ldots,\bh_n)\in U_1 \times \ldots \times U_n$, for any
$\delta>0$, for all $Q$ large enough, and for all non-zero $\bq\in \setZ_Q^m$,
$\|\bA(\bh_1,\ldots,\bh_n)\bq\|_2\ge Q^{-m/n-\delta}$. \hfill \QED
\end{lemma}

As far as DoF is concerned, the following lemma will be useful in
understanding the basis of our derivation. Its proof is provided in
\apref{ap.dof}.
\begin{lemma}\label{le.dof}
For a communication link described by \eqref{eq.basic}, where $\bA$ is a
matrix as defined in \leref{le.manifold}, then for almost all $(\bh_1,
\ldots,\bh_n)\in U_1 \times \ldots \times U_n$, the communication link based
on the resulting $\bA$ can provide a per-symbol DoF of $n/(m+n)$ and a total
DoF of $mn/(m+n)$.\hfill \QED
\end{lemma}

If the matrix $\bA$ represents a point to point MIMO system of $m$ transmit
antennas and $n$ receive antennas, then the achieved DoF $mn/(m+n)$ is smaller
than the maximum possible DoF $\min(m,n)$. However, if $n$ is the number of
receive antennas, and $m$ is the number of simultaneously transmitted symbols
using integer lattice, the total achieved DoF is $n$ when $m$ goes to
infinity. When using \leref{le.dof}, we will let $m \to\infty$ so that the gap
between the achieved DoF $mn/(m+n)$ based on a integer signaling and the
maximum DoF possible $\min(m,n)$ disappears.

\section{System model}

Consider a MIMO real Gaussian interference network with $K$ transmitters and
$J$ receivers. Suppose transmitter $k$ has $M_k$ antennas and receiver $j$ has
$N_j$ antennas. At each time, each transmitter, say transmitter $k$, sends a
vector signal $\bx_k\in\setR^{M_k}$. The channel from transmitter $k$ to
receiver $j$ is represented as a matrix
\begin{equation}
\bH_{j,k}\bydef [h_{j,k,r,t}]_{r=1,t=1}^{N_j,M_k}
\end{equation}
where $k\in\cK$, $j\in\cJ$, and $\bH_{j,k}\in \setR^{N_j\times M_k}$. It is
assumed that the channel is constant during all transmissions. Each transmit
antenna is subjected to an average power constraint $P$. The received signal
at receiver $j$ can be expressed as
\begin{equation}
\by_{j} = \sum_{k\in \cK} \bH_{j,k} \bx_{k} +
  \bnu_{j}, \quad \forall j\in \cJ
\end{equation}
where $\{\bnu_{j}|j\in \cJ\}$ is the set of independent Gaussian additive
noises with real, zero mean, independent, and unit variance entries. Let $\bH$
denote the \size{\sum_{j\in\cJ}{N_j}}x{\sum_{k\in\cK}{M_k}} block matrix,
whose $(j,k)$th block of size \size {N_j}x{M_k} is the matrix $\bH_{j,k}$. The
matrix $\bH$ includes all the channel coefficients.

In view of message demands at receivers, the introduced channel can specialize
to three known cases:
\begin{enumerate}
\item \emph{The $(K,J,[M_k],[N_j],[\cW_j])$ interference network with general
message demands}: where each receiver, for instance receiver $j$, requests an
arbitrary subsets of transmitted signals as $\cW_j=\{k\in\cK\st \text{receiver
$j$ requests } \bx_k \}$.
\item \emph{The single hop $(K,J,[M_k],[N_j])$ wireless X network}: where for
each pair $(j,k)\in\cJ\times\cK$, transmitter $k$ conveys an independent
message to receiver $j$.
\item \emph{The $K$-user interference channel}: where $J=K$ and signal
$\bx_k$, $\forall k\in\cK$, is just intended for receiver $k$. For this model,
we use the abbreviation $(K,[M_k],[N_j])$.
\end{enumerate}

In the case of $K$-user interference channel, the \emph{capacity region}
$\cC_{IC}(P,K,[M_k],[N_j],\bH)$ is defined in the usual sense: It contains
rate tuples $[R_k(P)]_{k=1}^K$ such that reliable transmission from
transmitter $k$ to receiver $k$ is possible at rate $R_k-\epsilon$, for any
$\epsilon>0$ and for all $k\in \cK$ simultaneously, under the given power
constraint $P$. Reliable transmissions mean that the probability of error can
be made arbitrarily small by increasing the encoding block length while
keeping the rates and power fixed.

A DoF vector $[d_k]_{k=1}^K$ is said to be \emph{achievable} if for any large
enough $P$, the rates $R_i=0.5\log(P) d_i$, $i=1,2,\ldots, K$, are
simultaneously achievable by all $K$ users, namely
\(0.5\log(P)\cdot [d_k]_{k=1}^K \in \cC_{IC}(P,K,[M_k],[N_j],\bH)\).
The \emph{DoF region} for a given interference channel $\bH$,
$\cD_{IC}(K,[M_k],[N_j],\bH)$, is the set of all achievable DoF vectors. The
DoF region $\cD_{IC}(K,[M_k],[N_j])$ is the largest possible region such that
$\cD_{IC}(K,[M_k],[N_j])\subset \cD_{IC}(K,[M_k],[N_j],\bH)$ for almost all
$\bH$ in the Lebesgue sense. The \emph{total DoF of the $K$-user interference
channel $\bH$} is defined as
\[
d_{IC}(K,[M_k],[N_j],\bH)=\max_{[d_k]_{k=1}^K\in \cD_{IC}(K,[M_k],[N_j],\bH)} \sum_{k=1}^K d_k.
\]
The \emph{total DoF $d_{IC}(K,[M_k],[N_j])$} is defined as the largest
possible real number $\mu$ such that for almost all (in the Lebesgue sense)
real channel matrices $\bH$ of size
\size{\sum_{j\in\cK}{N_j}}x{\sum_{k\in\cK}{M_k}},
$d_{IC}(K,[M_k],[N_j],\bH)\ge \mu$.

\remark The DoF region $\cD_{X}(K,J,[M_k],[N_j])$ for the single hop wireless
X network can be defined similarly as for the $K$-user interference channel
except in this case, any DoF point in the DoF region is a matrix of the form
$[d_{j,k}]_{j=1,k=1}^{J,K}$. Likewise, the DoF region
$\cD_{G}(K,J,[M_k],[N_j],[\cW_j])$ for interference network with general
message demand can be defined.

\section{Main Results} \label{sec.main}

The main results of our paper regarding achievable DoF regions are presented
below. The DoF region outer bound result will be presented in
\secref{sec.outer}.

\begin{theorem} \label{th.KNN}
$d_{IC}(K,[N],[N])=\frac{NK}{2}$. \hfill \QED
\end{theorem}
This result for constant coefficient channels has been obtained before in
\cite{mgmk09}. For time-varying channels, the same total DoF was established
in \cite{caja08}.

\begin{theorem} \label{th.KMN}
$d_{IC}(K,[M],[N])\ge \frac{MN}{M+N}K$. \hfill\QED
\end{theorem}
This result for constant coefficient channels has been obtained before in
\cite{ghmk09a}. For time-varying channels, the same total DoF was established
in \cite{goja10}.

\remark Our proofs for \thref{th.KNN} and \thref{th.KMN} are different from
those in \cite{mgmk09,ghmk09a} because antenna splitting is not employed. Our
scheme is more flexible in dealing with cases where the transmit messages do
not have the same DoF, in which case antenna splitting is not optimal.

\begin{theorem}\label{th.region.KMkNj}
The DoF region of a $(K,[M_k],[N_j])$ interference channel satisfies
$\cD_{IC}(K,[M_k],[N_j])\supset \cDin_{IC}$ where
\begin{equation}
\cDin_{IC}\bydef \{[d_k]_{k=1}^K\in \setR_+^{K\times 1}\st \frac{d_k}{N_k} + \max_{\hat
k\ne k} {\frac{d_{\hat k}}{M_{\hat k}}} \le 1, \forall k\in \cK\}.
\end{equation}
\end{theorem}

\begin{corollary}\label{th.region.KMN}
Setting all $M_K=M$ and $N_j=N$ in \thref{th.region.KMkNj}, the DoF region of
a $(K,[M],[N])$ interference channel satisfies $\cD_{IC}(K,[M],[N])\supset
\cDin_{IC}$ where
\begin{equation}
\cDin_{IC}\bydef \{[d_k]_{k=1}^K\in \setR_+^{K\times 1}\st M d_k + N\max_{\hat
k\ne k} d_{\hat k} \le MN, \forall k\in \cK\}.
\end{equation}
\end{corollary}

\begin{corollary}\label{th.region.KNN}
Let assume $M=N$ in \coref{th.region.KMN}. Employing the outer bound derived
in \cite{krwy12i}, the DoF region of a $(K,[N],[N])$ interference channel is
the following
\begin{equation}
\cD_{IC}(K,[N],[N])=\{[d_k]_{k=1}^K\in \setR_+^{K\times 1}\st d_k + \max_{\hat
k\ne k} d_{\hat k} \le N, \forall k\in \cK \}.
\end{equation}
\end{corollary}

\begin{theorem}\label{th.region.KJNNW}
The DoF region of a $(K,J,[N],[N],[\cW_j])$ interference network with general
message demand is
\begin{equation}
\cD_{G}(K,J,[N],[N],[\cW_j])\bydef \{[d_k]_{k=1}^K\in \setR_+^{K\times 1}\st
\sum_{k\in\cW_j}{d_k} + \max_{\hk\in\cW^c_j}{d_{\hk}} \le N, \forall j\in
\cJ\}.
\end{equation}
\end{theorem}

\begin{theorem}\label{th.region.KJMkNj}
The DoF region of a $(K,J,[M_k],[N_j])$ X network satisfies
$\cD_{X}(K,J,[M_k],[N_j])\supset \cDin_{X}$ where
\begin{equation}
\cDin_{X}\bydef \{[d_{j,k}]_{j=1,k=1}^{J,K}\in \setR_+^{K\times J}\st
\frac{1}{N_j}\sum_{k\in\cK}{d_{j,k}} +\\ \sum_{j\in\cJ, \hj\neq
j}{\max_{\hk\in\cK}{\frac{d_{\hj, \hk}}{M_\hk}}} \le 1, \forall j\in \cJ\}.
\end{equation}
\end{theorem}

\begin{corollary}\label{th.region.KJMN}
As a special case of \thref{th.region.KJMkNj}, the DoF region of a
$(K,J,[M],[N])$ X network channel satisfies $\cD_{X}(K,J,[M],[N])\supset
\cDin_{X}$ where
\begin{equation}
\cDin_{X}\bydef \{[d_{j,k}]_{j=1,k=1}^{J,K}\in \setR_+^{K\times J}\st M
\sum_{k\in\cK}{d_{j,k}} +\\ N\sum_{\hj\in\cJ, \hj\neq
j}{\max_{\hk\in\cK}{d_{\hj,\hk}}} \le MN, \forall j\in \cJ\}.
\end{equation}
\end{corollary}

\remark The same DoF regions as in \coref{th.region.KNN} and
\thref{th.region.KJNNW} for time-varying channel have been obtained before in
\cite{krwy12i} using vector alignment. It is interesting to note that the DoF
region is regardless of whether the channel is time-varying or constant. This
indicates that the DoF region for this channel is an inherent spatial property
of the channel that is separate from the time or frequency diversity, as has
been observed previously \cite{krwy12i,sggj12}.

\remark Employing the outer bound derived by \cite{caja09}, the achieved
region of \recol{th.region.KJMN} with the condition $M=N$ is tight in the
following cases:
\begin{enumerate}
\item The total number of receivers is $\cJ=2$.
\item $d_{j,k}=d_{j,\hk}$, for all $k, \hk\in\cK$ and for all $j\in\cJ$.
\end{enumerate}
If we set all $d_{j,k}=\frac{N}{K+J-1}$, then we obtain the total DoF
$\frac{KJN}{K+J-1}$. The same total DoF has been obtained in \cite{sggj12} for
time-varying channel. It is again notable that the total DoF does not depend
on the channel variability.

\remark If we set $M=1$ in \recol{th.region.KJMN}, we arrive at the
single-input multiple-output X network with $N$ antenna at all receivers. For
this model when $K>N$, we establish the total DoF $\frac{NKJ}{K+N(J-1)}$ by
fixing all $d_{j,k}=\frac{N}{K+N(J-1)}$ and employing the outer bound of
\cite{sggj12}. When $K\leq N$, beamforming and zeroforcing are sufficient to
achieve single-user outer bound $N$.

\remark The achievable DoF regions in
Theorems~\ref{th.region.KMkNj}--\ref{th.region.KJMkNj} are all of the
following type: i) there is one inequality for each receiver; ii) the
inequality is such that the total DoF of the useful messages, normalized by
the number of receive antennas, plus the sum, over the other receivers, of the
maximum interference DoF intended for each of these receivers, normalized by
the number of transmit antennas, is less than 1.

\remark \thref{th.KNN} follows from \thref{th.KMN} by setting $M=N$ and the
outer bound for $K$-user interference channel that has been obtained before in
\cite{caja08}. Moreover, \thref{th.KMN} follows from \recol{th.region.KMN}
when $d_k=MN/(M+N)$, $\forall k\in \cK$.

We conclude from the last remark that the only requirement to establish
\thref{th.KNN}--\ref{th.KMN} is proving \thref{th.region.KMkNj} (hence
\recol{th.region.KMN}). However, we will first prove the achievability of
\thref{th.KNN} in \secref{sec.KNN}, which serves to introduce the real
interference alignment scheme, joint antenna processing at the receivers, and
the performance analysis based on the results of simultaneous Diophantine
approximation on manifolds.

\section{total DoF of $(K,[N],[N])$ interference channel}\label{sec.KNN}

In this section, we examine our new achievability scheme on the $(K,[N],[N])$
interference channel. \thref{th.KNN} is then proved by employing the outer
bound in \cite{caja08}. Our scheme uses real interference alignment such that
the dimensions of interferences are aligned as much as possible, leaving more
dimensions for useful signals. The dimensions (also named directions) are
represented as real numbers that are rationally independent.

\part{encoding}: Transmitter $k$ sends a vector message $\bx_k =
{(x^1_k,\ldots,x^N_k)}^*$ where $x^t_k$, $\forall t\in\cN$ is the signal
emitted by antenna $t$ at transmitter $k$. The signal $x^t_k$ is generated
using transmit \emph{directions} in a set $\cT=\{T_i\in\setR\st 1\leq i\leq
D\}$ as $x^t_k = \bT \bs^t_k$ where $\bT \bydef (T_1,\ldots ,T_D)$, $\bs^t_k
\bydef {(s^t_{k1},\ldots , s^t_{kD})}^*$, and for all $1 \leq i \leq D$,
\begin{equation}\label{alpha}
s^t_{ki} \in \{\lambda q \st q\in\setZ, -Q\leq q\leq Q\}.
\end{equation}
The parameters $Q$ and $\lambda$ will be designed to satisfy the rate and
power constraints.

\begin{figure}[tbp]
\centering
\includegraphics[width=.7\linewidth]{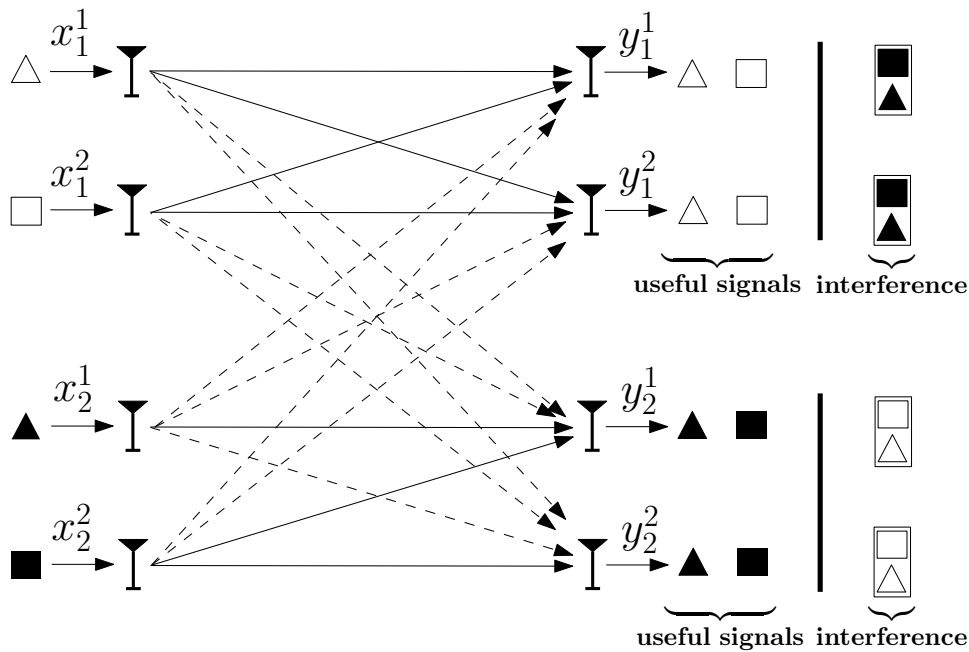}
\caption{2-user Gaussian interference channel with 2 antennas at each
transmitter receiver}
\label{fig.1}
\end{figure}

\part{Alignment Design}: We design transmit directions in such a way that at
any receiver antenna, each useful signal occupies a set of directions that are
rationally independent of interference directions.

To illustrate the idea, we use an example as depicted in \figref{fig.1}.
Messages $x^1_1$ and $x^2_1$ are shown by white triangle and square. In a
similar fashion, $x^1_2$ and $x^2_2$ are indicated with black triangle and
square. We are interested in the transmit directions such that at each
receiver antenna the interferences, for instance black triangle and square at
receiver~1, are aligned while the useful messages, white triangle and square,
occupy different set of directions.

\part{transmit directions}: Our scheme requires all directions of set $\cT$ to
be in the following form
\begin{equation}\label{eq.T}
T = \prod_{j\in \cK} \prod_{k\in\cK, k\ne j} \prod_{r\in\cN}\prod_{t\in\cN}
  \paren{h_{j,k,r,t}}^{\alpha_{j,k,r,t}}
\end{equation}
where
\(
0\leq \alpha_{j,k,r,t}\leq n-1,
\)
$\forall j\in \cK$, $k\in\cK$, $k\ne j$, $r\in\cN$, $t\in\cN$. It is easy to
see that the total number directions is
\begin{equation}\label{numbers}
D = n^{K(K-1)N^2}.
\end{equation}
We assume that directions in $\cT$ are indexed from 1 to $D$. The exact
indexing order is not important here. Note that in the single-input
single-output (SISO) case, the proposed transmission scheme coincides with the
scheme in \cite{mgmk09}.

\part{alignment analysis}: Our design proposes that at each antenna of
receiver $j$, $j\in\cK$, the set of messages $\{x_k^t \st k\in\cK, k\neq j,
t\in\cN\}$ are aligned. To verify, consider all $x_k^t$, $k\neq j$ that are
generated in directions of set $\cT$. These symbols are interpreted as the
interferences for receiver $j$. Let
\begin{equation}
D' = {(n+1)}^{K(K-1)N^2}.
\end{equation}
and define a set $\cT'=\{T'_i\in\setR \st 1\leq i\leq D'\}$ such that all
$T'_i$ are in from of $T$ as in \eqref{eq.T} but with a small change as
follows
\begin{equation}\label{eq.T'}
0\leq \alpha_{j,k,r,t}\leq n.
\end{equation}
Clearly, all $x_k^t$, $k\neq j$ arrive at antenna $r$ of receiver $j$ in the
directions of $\{\paren{h_{j,k,r,t}}T \st k\in\cK, k\neq j, t\in\cN,
T\in\cT\}$ which is a subset of $\cT'$.

This confirms that at each antenna of any receiver, all the interferences only
contain the directions from $\mathcal{T'}$. These interference directions can
be described by a vector
\(
\bT'  \bydef (T'_1,\ldots , T'_{D'}).
\)

\part{decoding scheme}: In this part, we first rewrite the received signals.
Then, we prove the achievability part of \thref{th.KNN} using \leref{le.dof}
based on joint antenna processing.

The received signal at receiver $j$ is represented by
\begin{equation}
\by_j = \underbrace{\mystrut{3.2ex}\bH_{j,j}\bx_j}_{\text{\small the useful signal}} + \underbrace{\sum_{k\in\cK,k\neq j}{\bH_{j,k}\bx_k}}_{\text{\small interference}}  + \bnu_{j}.
\end{equation}

Let us define
\begin{equation}
\bB \bydef \left(\begin{matrix}
  \bT &  \bzero  & \ldots & \bzero \cr
  \bzero   &  \bT & \ldots & \bzero \cr
  \vdots & \vdots &\ddots & \vdots\cr
  \bzero   &   \bzero &\ldots & \bT
  \end{matrix}\right), \quad
\bs_k \bydef \left(\begin{matrix}
   \bs^1_k \cr
   \bs^2_k \cr
   \vdots \cr
   \bs^N_k
   \end{matrix}\right), \quad
\bu_k\bydef\frac{\bs_k}{\lambda},
\end{equation}
such that $\bB$ is an $N\times ND$ matrix with $(N-1)D$ zeros at each row.
Using above definitions, $\by_j$ can be rewritten as
\begin{equation}
\by_j=\lambda\paren{ \bH_{j,j}\bB\bu_j+ \sum_{k\in\cK,k\neq
j}{\bH_{j,k}\bB\bu_k}} + \bnu_{j}.
\end{equation}
The elements of $\bu_k$ are integers between $-Q$ and $Q$, cf.~\eqref{alpha}.

We rewrite
\begin{equation}\label{reshape1}
\bH_{j,j}\bB\bu_j = \paren{\bH_{j,j}\otimes \bT} \bu_j = \\
				\left(\begin{matrix}
  h_{j,j,1,1}\bT &  h_{j,j,1,2}\bT  & \ldots & h_{j,j,1,N}\bT \cr
  h_{j,j,2,1}\bT   &  h_{j,j,2,2}\bT & \ldots & h_{j,j,2,N}\bT \cr
  \vdots & \vdots &\ddots & \vdots\cr
  h_{j,j,N,1}\bT & h_{j,j,N,2}\bT & \ldots & h_{j,j,N,N}\bT
  \end{matrix}\right)\bu_j \bydef
  \left(\begin{matrix}
  \bT^1_j\cr
  \bT^2_j\cr
  \vdots\cr
  \bT^N_j
  \end{matrix}\right)\bu_j
\end{equation}
where $\forall r\in\cN$, $\bT^r_j$ is the $r^{\text{th}}$ row of
$\bH_{j,j}\bB$. Also,
\begin{equation}\label{reshape2}
\sum_{k\in\cK,k\neq j}{\bH_{j,k}\bB\bu_k} = \sum_{k\in\cK,k\neq
j}{\paren{\bH_{j,k}\otimes\bT}\bu_k} = \\
				 \left(\begin{matrix}
  \sum_{k\in\cK, k\neq j}{\sum_{t\in \cN}{\paren{ h_{j,k,1,t}\bT\bu^t_k}}}\cr
  \sum_{k\in\cK, k\neq j}{\sum_{t\in \cN}{\paren{ h_{j,k,2,t}\bT\bu^t_k}}}\cr
  \vdots\cr
  \sum_{k\in\cK, k\neq j}{\sum_{t\in\cN}{\paren{ h_{j,k,N,t}\bT\bu^t_k}}}
  \end{matrix}\right) \overset{(a)}{=}
  \left(\begin{matrix}
  \bT'\bu'^1_j\cr
  \bT'\bu'^2_j\cr
  \vdots\cr
  \bT'\bu'^N_j
  \end{matrix}\right)
\end{equation}
where $\forall r\in\cN$, $\bu'^r_j$ is a column vector with $D'$ integer
elements (some of the entries are zero), and $(a)$ follows since the set
$\cT'$ contains all directions of the form $\paren{h_{j,k,r,t}}T$ where $k\neq
j$; cf.~the definition of $\cT'$.

Considering (\ref{reshape1}) and (\ref{reshape2}), we are able to equivalently
denote $\by_j$ as
\begin{equation}\label{zeros}
\by_j = \lambda \left(\begin{matrix}
   \bT^1_j &\bT' &  \bzero  & \ldots & \bzero \cr
  \bT^2_j &\bzero   &  \bT' & \ldots & \bzero \cr
  \vdots &\vdots & \vdots &\ddots & \vdots\cr
  \bT^N_j &\bzero   &   \bzero &\ldots & \bT'
  \end{matrix}\right)
  \left(\begin{matrix}
  \bu_j\cr
  \bu'^1_j\cr
  \vdots\cr
  \bu'^N_j
  \end{matrix}\right)
  + \bnu_{j}.
\end{equation}
It should be pointed out $\bT^r_j$ represents the useful directions at antenna
$r$ of receiver $j$. The elements in $\bT'$ represent the interference
directions, which is common to all antennas at all receivers.

We finally left multiply $\by_j$ by an $N\times N$ weighting matrix
\begin{equation}\label{numbers3}
\bW= \left(\begin{matrix}
  1 &  \gamma_{12}& \ldots & \gamma_{1N}  \cr
  \gamma_{21}  & 1 & \ldots & \gamma_{2N} \cr
  \vdots & \vdots & \ddots & \vdots \cr
  \gamma_{N1}& \gamma_{N2} &\ldots & 1
  \end{matrix}\right)
\end{equation}
such that all indexed $\gamma$ can be chosen randomly, and independently from
any continuous distribution, say, uniformly from the interval $[\frac{1}{2},
1]$. This process causes the zeros in (\ref{zeros}) to be filled by non-zero
directions.

After multiplying $\bW$, the noiseless received constellation belongs to a
lattice generated by the $N\times N(D+D')$ matrix
\begin{equation}\label{A}
 \bA = \bW \left(\begin{matrix}
   \bT^1_j &\bT' &  \bzero  & \ldots & \bzero \cr
  \bT^2_j &\bzero   &  \bT' & \ldots & \bzero \cr
  \vdots &\vdots & \vdots &\ddots & \vdots\cr
  \bT^N_j &\bzero   &   \bzero &\ldots & \bT'
  \end{matrix}\right).
\end{equation}

The above matrix has a significant property that allows us to use
\leref{le.manifold}. More precisely, \leref{le.manifold} requires each row of
$\bA$ to be a non-degenerate map from a subset of channel coefficients to
$\setR^{N(D+D')}$. The non-degeneracy is established because
(cf.~\apref{ap.nondegen}):
\begin{enumerate}
\item all elements of $\bT'$ and $\bT^t_j$, $\forall t\in\cN$ are analytic
functions of the channel coefficients;
\item all the directions in $\bT'$ and $\bT^t_j$, $\forall t\in\cN$ together
with 1 are linearly independent over $\setR$ ;
\item all indexed $\gamma$ in $\bW$ have been chosen randomly and
independently.
\end{enumerate}
Since $ {\|\bq\|}_{\infty}\le (K-1)NQ$, for any $\delta>0$ and large enough
$Q$, the distance between any two points of the received constellation
(without considering noise) is lower bounded via \leref{le.manifold} by
\begin{equation}\label{result1}
\lambda \bigl(\,2(K-1)NQ\,\bigr)^{-(D+D')-\delta}.
\end{equation}

We now focus our attention on the design of $\lambda$ and $Q$ to complete the
coding scheme. The parameter $\lambda$ controls the input power of transmitter
antennas. The average power of antenna $t$ at transmitter $k$ is computed as
\begin{equation}
P = E[({x_k^t})^2] = E[{(\bT\bs_k^t)}^2] = \sum_{i=1}^D{{T_i}^2E[({s_{ki}^t})^2]}
 \leq \lambda^2 Q^2 \sum_{i=1}^D{{T_i}^2}\bydef \lambda^2 Q^2 \nu^2
\end{equation}
where the inequality follows from equation \eqref{alpha} and $\nu^2 \bydef
\sum_{i=1}^D{{T_i}^2} $. Thus, the only requirement to satisfy the power
constraint is $\lambda \leq \frac{P^{\frac{1}{2}}}{Q \nu}$. It is sufficient
to choose
\begin{equation}\label{lambda}
\lambda=\frac{\zeta P^{\frac{1}{2}}}{Q},
\end{equation}
where $\zeta = \frac{1}{\nu}$.

Let $P_0=\lambda Q=P/\nu^2$. By \leref{le.dof}, each symbol $s^t_{ki}$ can
achieve a rate of $d_0 \log(P_0)$ for large $P_0$, where
$d_0=N/[N+N(D+D']=1/(1+D+D')$. Since there are totally $ND$ useful symbols
from each transmitter, the total achievable rate, as normalized by $\log(P_0)$
for each transmitter is
\begin{equation}
\frac{ND}{D+D'+1} =\frac{Nn^{K(K-1)N^2}}{n^{K(K-1)N^2}+{(n+1)}^{K(K-1)N^2}+1}
\end{equation}
and as $n$ increases, it converges to $\frac{N}{2}$. Since $P$ and $P_0$ are
different by a multiplication factor $\nu^2$, when the rate is normalized by
$\log(P)$ instead, as required in the definition of DoF, the same limit of
$N/2$ will result as the per user DoF, as $P\to\infty$. The total DoF of the
$K$ users is therefore $NK/2$, which meets the outer bound \cite{caja08}. This
finishes the proof of the achievability of the total DoF. When combined with
the corresponding outer bound, the theorem is proved.

\section{$K$-user interference channel and inner bound on DoF
region}\label{sec.region.KMN}

For simplicity, we will first prove \recol{th.region.KMN} in this section.
Then utilizing the presented proof, \thref{th.region.KMkNj} will be
established.

Consider a $(K,[M],[N])$ MIMO interference channel. We prove that for any
$[d_k]_{k=1}^{K}\in\cDin_{IC_{1}}$, $[d_k]_{k=1}^{K}$ is achievable.

Assume that it is possible to find an integer $\rho$ such that $\forall
k\in\cK$, ${\bar{d}}_k=\rho \frac{d_k}{M}$ is a non-negative integer. The
signal $x_k^t$ is divided into ${\bar{d}}_k$ streams. For stream $l$, $l\in
\{1,\ldots,\displaystyle\max_{k\in\cK}{\bar{d}_k}\}$, we use directions
$\{T_{l1},\ldots, T_{lD}\}$ of the following form
\begin{equation}\label{numbers2}
T_l = \prod_{j\in \cK} \prod_{k\in\cK, k\ne j} \prod_{r\in\cM}\prod_{t\in\cN}
  \paren{h_{j,k,r,t}\delta_l}^{\alpha_{j,k,r,t}}
\end{equation}
where $0 \leq \alpha_{j,k,r,t} \leq n-1$ and $\delta_l$ is a design parameter
that is chosen randomly, independently, and uniformly from the interval
$[\frac{1}{2},1]$. Let $\bT_l \bydef (T_{l1}, \ldots, T_{lD})$. Note that, at
any antenna of transmitter $k$, the constants $\{\delta_l\}$ cause the streams
to be placed in $\bar{d}_k$ different sets of directions. Indeed the constants
$\{\delta_l\}$ play the role analogous to the base vectors $\mathbf{w}_i$ in
\cite{krwy12i}. The alignment scheme is the same as before, considering the
fact that at each antenna of receiver $j$, the useful streams occupy
$M\bar{d}_j$ separate sets of directions. The interferences are also aligned
at most in $\displaystyle\max_{k\in\cK, k\neq j}{\bar{d}_k}$ sets of
directions independent from useful directions.

By design, $x_k^t$ is emitted in the following form
\begin{equation}
x_k^t = \sum_{l=1}^{\bar{d}_k}{\delta_l}\sum_{i=1}^{D}{T_{li}s_{kli}^t} =
\bT_k\bs^t_k
\end{equation}
where
\begin{equation}
\bT_k\bydef(\delta_1 \bT_{1},\ldots,\delta_{\bar{d}_k}\bT_{\delta_{\bar{d}_k}}),
\quad
\bs^t_k \bydef {(s^t_{k11},\ldots , s^t_{k\bar{d}_kD})}^*,
\end{equation}
and all $s_{kli}^t$ belong to the set defined in (\ref{alpha}).

Pursuing the same steps of the previous section for receiver $j$, $\bB$
becomes an $M\times MD\bar{d}_j$ matrix as
\begin{equation}
 \left(\begin{matrix}
  \bT_j &  \bzero  & \ldots & \bzero \cr
  \bzero   &  \bT_j & \ldots & \bzero \cr
  \vdots & \vdots &\ddots & \vdots\cr
  \bzero   &   \bzero &\ldots & \bT_j
  \end{matrix}\right)
\end{equation}
and $\bA$ will have $N$ rows and $MD{\bar{d}}_j+ND'\displaystyle\max_{k\in\cK,
k\neq j}{{\bar{d}}_k}$ columns. To be more precise, matrix $\bA$ has the same
form as \eqref{A} noting that $\bT^r_j$ and $\bT'$ are now vectors with
$MD\bar{d}_j$ and $D'\displaystyle\max_{k\in\cK, k\neq j}{\bar{d}_k}$
elements, respectively.

\remark As it has been proved in the previous section, the dimensions of
matrix $\bA$ inherits two characteristics as follows:
\begin{enumerate}
\item The number of columns is the number of all available directions at the
receiver.
\item For large $n$, the number of rows over the number of columns specifies
the achievable DoF per direction.
\end{enumerate}

Let $G_j$ denote the number of columns of $\bA$. For any DoF points in
$\cDin_{IC}$ satisfying \recol{th.region.KMN}, we have
\begin{equation}
G_j = MD{\bar{d}}_j  + ND'{\displaystyle\max_{k\in\cK, k\neq j}{{\bar{d}}_k}}
\leq \frac{\rho}{M}NMD' = \rho ND'
\end{equation}
and as $n$ increases, the DoF of the signal $\bx_j$ intended for receiver $j$,
$\forall j\in\cK$ is at least
\begin{equation} \label{DoF}
\lim_{n\to\infty}{MD\bar{d}_j\frac{N}{G_j}} \ge
\lim_{n\to\infty}{MD\bar{d}_j\frac{N}{\rho ND'}}= \\
\lim_{n\to\infty}\frac{M}{\rho}\frac{\bar{d}_jn^{K(K-1)N^2}}{{(n+1)}^{K(K-1)N^2}}
= \frac{M}{\rho}\bar{d}_j = d_j
\end{equation}
where $\frac{N}{\rho ND'}$ is the DoF per direction for large $D'$. This
proves \coref{th.region.KMN}.

As a special case, it is easy to see when all $d_k$ are equal, the total
achievable DoF is $\frac{MN}{M+N}K$. Moreover, when $M=N$, the achievable DoF
region is tight, cf.~\reref{outer.KNN}.

To establish \thref{th.region.KMkNj}, we follow the proof of
\recol{th.region.KMN} with a small change in assumption, which is
${\bar{d}}_k=\rho \frac{d_k}{M_k}$. As a result, $\bA$ becomes $N_j$ by
$M_kD{\bar{d}}_j+N_jD'\displaystyle\max_{k\in\cK, k\neq j}{{\bar{d}}_k}$
matrix. Therefore, for any DoF points in $\cDin_{IC}$ satisfying
\thref{th.region.KMkNj}, we have
\begin{equation}
G_j = M_KD{\bar{d}}_j  +
  N_jD'{\displaystyle\max_{k\in\cK, k\neq j}{{\bar{d}}_k}} \leq \rho N_jD'
\end{equation}
and the DoF of signal $x_j$ is finally obtained as
\begin{equation}
\lim_{n\to\infty}{M_kD\bar{d}_j\frac{N_j}{\rho N_jD'}}= \\
\lim_{n\to\infty}d_j\frac{n^{K(K-1)N^2}}{{(n+1)}^{K(K-1)N^2}} = d_j.
\end{equation}

\section{Interference network with general message demands} Consider a
$(K,J,[N],[N],[\cW_j])$ single hop interference network with general message
demand. Transmitter $k$ emits independent message $\bx_k$, and receiver $j$
requests an arbitrary subset of messages denoted by $\cW_j$. We follow the
same definitions and steps of \secref{sec.region.KMN} considering stream $l$,
uses directions of the following form
\begin{equation}
T_l = \prod_{j\in \cJ} \prod_{k\in\cW^c_j} \prod_{r\in\cN}\prod_{t\in\cN}
  \paren{h_{j,k,r,t}\delta_l}^{\alpha_{j,k,r,t}}
\end{equation}
where $0 \leq \alpha_{j,k,r,t} \leq n-1$, $\cW^c_j \bydef \{k\in\cK \st
k\notin\cW_j\}$, and $\delta_l$ is a design parameter chosen as before. Notice
that the directions has been designed in such a manner that at any receiver,
for example receiver $j$, while the useful signal subspace is separated from
the interference subspace, all interferences caused by $\bx_k$, $k\in\cW_j$
are aligned. As a result, matrix $\bA$ at receiver $j$ will have $N$ rows and
$\displaystyle
ND\sum_{k\in\cW_j}{\bar{d}_k}+ND'\max_{\hk\in\cW^c_j}{\bar{d}_{\hk}}$ columns.
Thus, for any DoF point in $\cDin_{G}$ satisfying \thref{th.region.KJNNW},
$G_j$ is upper bounded by $\rho ND'$ and $d_k$, $k\in\cW_j$, is achieved
similar to \eqref{DoF}. The proof of the converse is the same as in
\cite{krwy12i}.

\section{wireless X networks} Consider a $(K,J,[M],[N])$ Gaussian X network.
For each pair $(j,k)\in \cJ\times\cK$, transmitter $k$ sends an $M\times 1$
vector message $\bx_{j,k} = {(x^1_{j,k},\ldots ,x^M_{j,k})}^*$ to receiver
$j$. Consequently, the signal emitted by transmitter $k$ is in the following
form
\begin{equation}
\bx_k = \sum_{j\in\cJ}{\bx_{j,k}}.
\end{equation}

We assume that it is possible to find an integer $\rho$ such that for all
$j\in\cJ$ and all $k\in\cK$, $\bar{d}_{j,k}=\rho \frac{d_{j,k}}{M}$ is a
non-negative integer. Message $x^t_{j,k}$ is divided into $\bar{d}_{j,k}$
streams such that each stream, say stream
$l\in\{1,\ldots,\displaystyle\max_{k\in\cK}{\bar{d}_{j,k}}\}$, uses directions
in set $\cT_{j,l}=\{T_{j,l,i}\in\setR\st 1\leq i\leq D\}$. All $T_{j,l,i}$ are
generated in the following form
\begin{equation}\label{Tjkl}
T_{j,l} = \prod_{\hj\in \cJ, \hj\neq j} \prod_{\hk\in\cK} \prod_{\substack{\\ \hr\in\cN}} \prod_{\hht\in\cM}\paren{h_{\hj,\hk,\hr,\hht}\delta_{j,l}}^{\alpha_{\hj,\hk,j,\hr,\hht,l}}
\end{equation}
where $0 \leq \alpha_{\hj,\hk,j,\hr,\hht,l} \leq n-1$ and $\delta_{j,l}$ is a
design parameter that is chosen randomly, independently, and uniformly from
the interval $[\frac{1}{2},1]$. Define $\bT_{j,l} \bydef (T_{j,l,1}, \ldots,
T_{j,l,D})$. The signal $x^t_{j,k}$ is generated as
\begin{equation}\label{xtjk}
x^t_{j,k} =\sum_{l=1}^{\bar{d}_{j,k}}{\delta_{j,l}}\sum_{i=1}^{D}{T_{j,l,i}s_{j,k,l,i}^t} = \bU_{j,k}\bs_{j,k}^t
\end{equation}
where
\begin{equation}
\bU_{j,k} = (\delta_{j,1}\bT_{j,1},\ldots, \delta_{j,\bar{d}_{j,k}}\bT_{j,\bar{d}_{j,k}}),
\end{equation}
\begin{equation}
\bs_{j,k}^t = {(s_{j,k,1,1}^t,\ldots,s_{j,k,\bar{d}_{j,k},D}^t)}^*,
\end{equation}
and all $s_{j,k,l,i}^t$ are members of the set in \eqref{alpha}.

\part{alignment design}: Suppose we are at receiver $j$. The design of
transmit directions guarantees that at any antenna of receiver $j$, the useful
signals are placed in $K$ separate sets of directions. Each set has
$D\bar{d}_{j,k}$, $k\in\cK$ directions. The interferences are also put in
$J-1$ different sets of directions, each containing all signals intended for
receiver $\hj$, $\hj\in\cJ$, $\hj\neq j$ with at most
$D'\displaystyle\max_{k\in\cK}{\bar{d}_{\hj,k}}$ directions.

\begin{figure}[tbp]
\centering
\includegraphics[width=.7\linewidth]{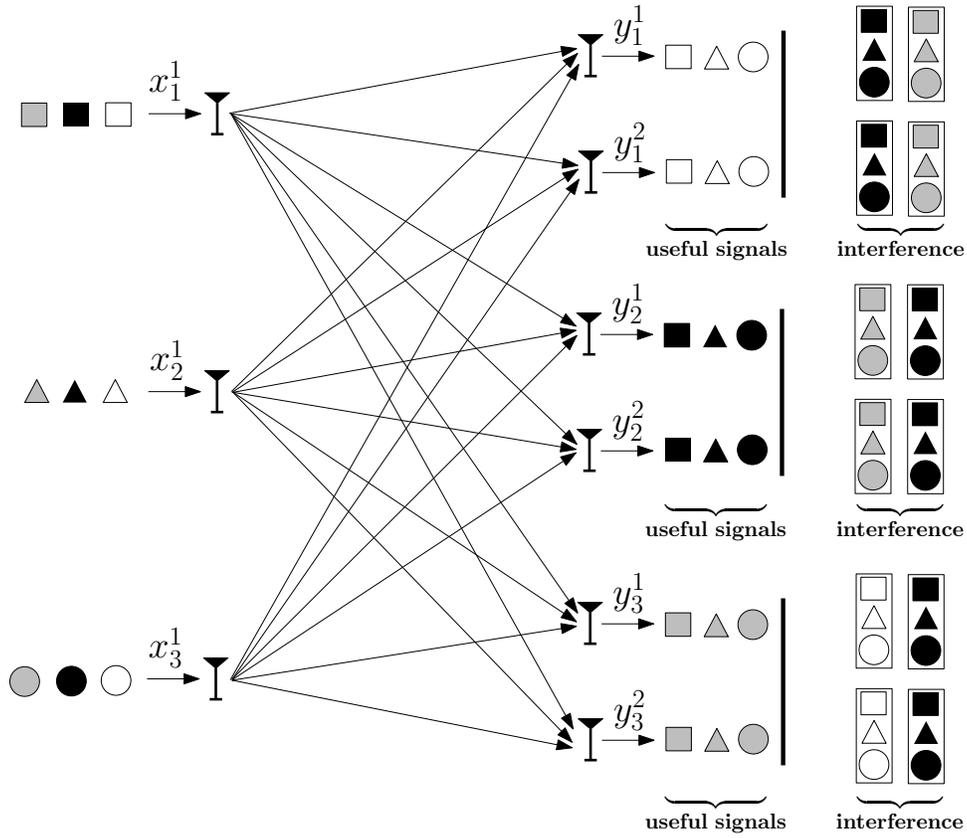}
\caption{($3\times 3, 1, 2$) Gaussian X network channel}
\label{fig.2}
\end{figure}

Let us explain the above mentioned argument for a $(3,3,[1],[2])$ Gaussian X
network. This system is depicted in \figref{fig.2}. Each transmitter conveys
an independent message to each receiver. We have assumed that white square,
triangle, and circle are the useful signals for the first receiver. Similarly,
black and gray nodes show the signals intended for receiver $2$ and $3$,
respectively. The transmission scheme is such that at any antenna of receiver
1:
\begin{itemize}
  \item The interferences, black square triangle and circle, are aligned. The gray signals are also aligned.
  \item The useful signals, white square triangle and circle, are not aligned.
\end{itemize}
Hence, at each receive antenna of first user, we have the sum of five terms
made by three useful signals and two sets of aligned signals. The set of
directions used for each term is separate from others in sense of rational
independence. A similar statement is also valid for other receivers. We prove
\thref{th.region.KJMN} provided that the described alignment scheme is
successful.

\part{alignment verification}: The proposed transmit directions guarantee that
the interferences created by messages intended for the same receiver are
aligned at all other receivers. To see this, let us define
$\cT'_{j,l}=\{T'_{j,l,i}\in\setR \st 1\leq i\leq D'\}$ such that all
$T'_{j,l,i}$ are in the form of \eqref{Tjkl} but with $0 \leq
\alpha_{\hj,\hk,j,\hr,\hht,l} \leq n$. We use $\bT'_{j,l}$ to denote vector
$(T'_{j,l,1},\ldots,T'_{j,l,D'})$. According to \eqref{xtjk}, the
$l^{\text{th}}$ stream of message $x_{j,k}^t$ is transmitted in directions of
the form $\delta_{j,l}T_{j,l}$. This stream arrives at antenna $r$ of receiver
$\hj$, $\hj\neq j$, in directions of the form
$\paren{h_{\hj,k,r,t}\delta_{j,l}}T_{j,l}$, which are obviously in set
$\cT'_{j,l}$. Since $\cT'_{j,l}$ does not depend on indices $k$ and $r$,
cf.~\eqref{Tjkl}, at any antenna of receiver $\hj$, $\hj\neq j$, all
directions created by the streams intended for receiver $j$ are subset of
$\cT'_{j,l}$, $\forall
l\in\{1,\ldots,\displaystyle\max_{k\in\cK}{\bar{d}_{j,k}}\}$ and occupy at
most $D'\displaystyle\max_{k\in\cK}{\bar{d}_{j,k}}$ dimensions. We denote
these directions as a vector
$\bT'_{j}\bydef(\bT'_{j,1},\ldots,\bT'_{j,\max_{k\in\cK}{\bar{d}_{j,k}}})$.

\part{decoding scheme}: The received signal at receiver $j$ can be divided
into two parts, the useful signals and interference, of the following form
\begin{equation}
\by_{j} =\sum_{k\in\cK}{\bH_{j,k}\bx_{j,k}} + \sum_{k\in\cK}\sum_{\hj\in\cJ,\hj\neq j}{\bH_{j,k}\bx_{\hj,k}}+\bnu.
\end{equation}

For notational convenience, let $\bs_{j,k}\bydef
{(\bs^1_{j,k},\ldots,\bs^M_{j,k})}^*$ and
$\bu_j\bydef\frac{1}{\lambda}{(\bs_{j,1},\ldots,\bs_{j,K})}^*$ with integer
elements between $-Q$ and $Q$. Then, we can rewrite the useful signals as
follows
\begin{align}
\sum_{k\in\cK}{\bH_{j,k}\bx_{j,k}}
=\sum_{k\in\cK}{\bH_{j,k}\left(\begin{matrix}
  x_{j,k}^1\cr
  x_{j,k}^2\cr
  \vdots\cr
  x_{j,k}^M
  \end{matrix}\right)}
 & \overset{(b)}
  =\sum_{k\in\cK}{\left(\begin{matrix}
  h_{j,k,1,1}\bU_{j,k} &  h_{j,k,1,2}\bU_{j,k} & \ldots & h_{j,k,1,N}\bU_{j,k} \cr
  h_{j,k,2,1}\bU_{j,k}   &  h_{j,k,2,2}\bU_{j,k} & \ldots & h_{j,k,2,N}\bU_{j,k} \cr
  \vdots & \vdots &\ddots & \vdots\cr
  h_{j,k,N,1}\bU_{j,k} & h_{j,k,N,2}\bU_{j,k} & \ldots & h_{j,k,N,N}\bU_{j,k}
  \end{matrix}\right)\bs_{j,k}} \\
   & \bydef \sum_{k\in\cK}{\left(\begin{matrix}
  \bU_{j,k}^1\cr
  \bU_{j,k}^2\cr
  \vdots\cr
  \bU_{j,k}^N
  \end{matrix}\right)\bs_{j,k}}
 = \lambda \left(\begin{matrix}
  \bU_{j,1}^1 &  \bU_{j,2}^1 & \ldots & \bU_{j,K}^1 \cr
  \bU_{j,1}^2   & \bU_{j,2}^2 & \ldots & \bU_{j,K}^2 \cr
  \vdots & \vdots &\ddots & \vdots\cr
  \bU_{j,1}^N  & \bU_{j,2}^N  & \ldots & \bU_{j,K}^N
  \end{matrix}\right)\bu_j\label{Uj}
\end{align}
where $\bU_{j,k}^r \bydef (h_{j,k,r,1}\bU_{j,k}, h_{j,j,r,2}\bU_{j,k}, \ldots
, h_{j,j,r,N}\bU_{j,k})$, $\forall j\in\cJ$, $k\in\cK$, $r\in\cN$. Using the
definition in \eqref{xtjk}, $(b)$ follows. We take into account that none of
$\cT'_{\hj,l}$, $\hj\neq j$, contains generators
$\{\paren{h_{j,k,r,t}\delta_{j,l}}\st k\in\cK, r\in\cN, t\in\cM\}$. Hence, the
directions in all $\bU_{j,k}^r$ and $\bT'_{\hj}$, $\hj\neq j$ are rationally
independent.

The interference part can be written as
\begin{align}
\sum_{k\in\cK}\sum_{\hj\in\cJ,\hj\neq j}{\bH_{j,k}\bx_{\hj,k}} {\ }
& = \sum_{\hj\in\cJ,\hj\neq j}\sum_{k\in\cK}{\bH_{j,k}\left(\begin{matrix}
  x_{\hj,k}^1\cr
  x_{\hj,k}^2\cr
  \vdots\cr
  x_{\hj,k}^M
  \end{matrix}\right)} \overset{(c)}{=}
 \sum_{\hj\in\cJ,\hj\neq j}\sum_{k\in\cK}{\bH_{j,k}\left(\begin{matrix}
  \bU_{\hj,k} \bs_{\hj,k}^1\cr
  \bU_{\hj,k} \bs_{\hj,k}^2\cr
  \vdots\cr
  \bU_{\hj,k} \bs_{\hj,k}^M
  \end{matrix}\right)} \nonumber \\
&= \sum_{\hj\in\cJ,\hj\neq j}{\left(\begin{matrix}
  \sum_{k\in\cK}\sum_{t\in\cM}{\paren{h_{j,k,1,t}\bU_{\hj,k}\bs_{\hj,k}^t}}\cr
  \sum_{k\in\cK}\sum_{t\in\cM}{\paren{h_{j,k,2,t}\bU_{\hj,k}\bs_{\hj,k}^t}}\cr
  \vdots\cr
  \sum_{k\in\cK}\sum_{t\in\cM}{\paren{h_{j,k,N,t}\bU_{\hj,k}\bs_{\hj,k}^t}}
  \end{matrix}\right)} \overset{(d)}{=}
    \sum_{\hj\in\cJ,\hj\neq j}{\lambda\left(\begin{matrix}
  \bT'_{\hj}\bu'^1_{\hj}\cr
  \bT'_{\hj}\bu'^2_{\hj}\cr
  \vdots\cr
  \bT'_{\hj}\bu'^N_{\hj}
  \end{matrix}\right)}\label{Ihj}
\end{align}
where for all $r\in\cN$, $\bu'^r_{\hj}$ is a column vector with integer
elements. Equivalence relation $(c)$ follows from \eqref{xtjk}. The equality
$(d)$ is due to alignment by our design. It is convenient to represent
equation \eqref{Ihj} as
\begin{equation}\label{Ic}
 \lambda\left(\begin{matrix}
  {\bI_j}{\bz}_{j}^1\cr
  {\bI_j}{\bz}_{j}^2\cr
  \vdots\cr
  {\bI_j}{\bz}_{j}^N
  \end{matrix}\right)
\end{equation}
where $\bI_j \bydef (\bT'_1,\ldots ,\bT'_{j-1},\bT'_{j+1},\ldots ,\bT'_J)$ and
$\bz_j^r \bydef (\bu'^r_1,\ldots ,\bu'^r_{j-1},\bu'^r_{j+1},\ldots ,\bu'^r_J)
$ for all $t\in\cN$.

Using \eqref{Uj} and \eqref{Ic}, received signal $\by_j$ is represented by
\begin{equation}
  \lambda \left(\begin{matrix}
  \bU^1_{j,1} & \bU^1_{j,2} & \ldots & \bU^1_{j,K}  &\bI_{j} &  \bzero  & \ldots & \bzero \cr
  \bU^2_{j,1} & \bU^2_{j,2} & \ldots & \bU^2_{j,K} &\bzero   &  \bI_{j}  & \ldots & \bzero \cr
  \vdots & \vdots & \ddots & \vdots & \vdots &\vdots &\ddots & \vdots\cr
  \bU^N_{j,1} & \bU^N_{j,2} & \ldots & \bU^N_{j,K} &\bzero   &   \bzero &\ldots & \bI_{j}
  \end{matrix}\right)
  \left(\begin{matrix}
  \bu_j\cr
  \bz^1_j\cr
  \vdots\cr
  \bz^N_j
  \end{matrix}\right)
  + \bnu_{j}.
\end{equation}

Analogous to achievability proof of \thref{th.KNN}, we left multiply $\by_j$
by an $N\times N$ weighting matrix. Then, $\bA$ in \eqref{A} becomes an
$N\times (MD\sum_{k\in\cK}{\bar{d}_{j,k}}+ND'\sum_{\hj\in\cJ,\hj\neq
j}{\max_{k\in\cK}{\bar{d}_{\hj,k}}})$ matrix such that the non-degeneracy
conditions is satisfied.

For any DoF point in $\cDin_{XC_{1}}$ that satisfies \thref{th.region.KJMN},
the total directions $G_j$ of the useful signals and the interferences at
receiver $j$ is
\begin{equation}
G_j = MD\sum_{k\in\cK}{\bar{d}_{j,k}}+
  ND'\sum_{\hj\in\cJ,\hj\neq j}{\max_{k\in\cK}{\bar{d}_{\hj,k}}}
\leq \rho ND'.
\end{equation}
Thus, as $n$ increases, the DoF of $\bx_{j,k}$, $\forall j\in\cJ$, $k\in\cK$,
is at least
\begin{equation}
\lim_{n\to\infty}{MD\bar{d}_{j,k}\frac{N}{\rho ND'}}= \\
\lim_{n\to\infty}\frac{M}{\rho}\frac{\bar{d}_{j,k}n^{K(K-1)N^2}}{{(n+1)}^{K(K-1)N^2}}
= \frac{M}{\rho}\bar{d}_{j,k} = d_{j,k},
\end{equation}
which establishes \thref{th.region.KJMN}.

The provided scheme for the $(K,J,[M],[N])$ Gaussian X network channel can be
applied to a more general case where each transmitter$/$receiver has an
arbitrary number of antennas. Let us assume that transmitter $k$ has $M_k$
antennas and receiver $j$ has $N_j$ antennas. To prove
\thref{th.region.KJMkNj}, we follow the same procedure of this section for
receiver $j$ considering the integer $\rho$ is changed such that
$\bar{d}_{j,k}=\rho \frac{d_{j,k}}{M_k}$, $\forall k\in\cK$, $j\in\cJ$.
Accordingly, $\bA$ becomes an $N\times (D\sum_{k\in\cK}{M_k \bar{d}_{j,k}}+N_j
D'\sum_{\hj\in\cJ,\hj\neq j}{\max_{k\in\cK}{\bar{d}_{\hj,k}}})$ matrix. Hence,
the total number of useful and interference directions at receiver $j$ is
\begin{equation}
G_j =  D\sum_{k\in\cK}{M_k \bar{d}_{j,k}}+N_j D'\sum_{\hj\in\cJ,\hj\neq j}{\max_{k\in\cK}{\bar{d}_{\hj,k}}}
\end{equation}
and $G_j\leq \rho N_j D'$ for any DoF point in $\cDin_{XC_{2}}$ satisfying
\thref{th.region.KJMkNj}. As a result, for large enough $n$, the DoF of signal
$\bx_{j,k}$ is attained as
\begin{equation}
\lim_{n\to\infty}{M_k D\bar{d}_{j,k}\frac{N_j}{\rho N_j D'}}= \\
\lim_{n\to\infty}\frac{M_k}{\rho}\frac{\bar{d}_{j,k}n^{K(K-1)N^2}}{{(n+1)}^{K(K-1)N^2}}
= \frac{M_k}{\rho}\bar{d}_{j,k} = d_{j,k}
\end{equation}
for all $j\in\cJ$ and $k\in\cK$. This completes the proof.

\section{outer bound discussion}\label{sec.outer}

Although our focus in this paper is on the new receive antenna joint
processing, we present a brief discussion on existing outer bounds of
interference networks. Note that all outer bounds are general as it applies to
interference networks regardless of whether the channel coefficients are time
varying or constant. We also present a new outer bound on the DoF region based
on a known technique of transmitter grouping.

Ghasemi et al. in \cite{ghmk10c} show that the total DoF of $(K,[M],[N])$ MIMO
Gaussian interference channel is outer bounded by $K\frac{MN}{M+N}$ when
$K\geq\frac{M+N}{\text{gcd}(M,N)}$. To establish this result, first consider
an $(L,[M],[N])$ MIMO interference channel where $L\leq K$. For this scenario,
the $L$ users are divided into two arbitrary disjoint sets of size $L_1$ and
$L_2$ such that $L=L_1+L_2$. The full cooperation among transmitters in each
set is assumed and similarly for each set of receivers. Accordingly, the
$2$-user MIMO interference channel with $L_1M$, $L_2M$ antenna at transmitters
and $L_1N$, $L_2N$ antennas at receivers is obtained. Using the DoF region of
$2$-user MIMO interference channel \cite{jafa07}, the DoF is finally outer
bounded.

It is also shown that for $K\leq \frac{\max{(M,N)}}{\min{(M,N)}}+1$, the total
DoF outer bound is $\min{(M,N)}\min{(K,\frac{\max{(M,N)}}{\min{(M,N)}})}$.
However, the DoF characterization for the remaining region
$\lfloor\frac{\max{(M,N)}}{\min{(M,N)}}\rfloor+1<K<\frac{M+N}{\text{gcd}(M,N)}$
has not been established due to the complexity of convex optimizations over
integers. To understand the origin of this problem, we next examine the
mentioned scheme when $L_2M$ has the minimum difference from $L_1N$ and we
extend the result to obtain an outer bound on the DoF region.

The key to establishing the outer bound on $(K,[M],[N])$ interference channel
is to consider a set of $g$ receivers as a group. For this receiver set, the
corresponding transmitters emitting useful signals are assumed to be
cooperative as one set. Hence, the rest of transmitters only create
interference. We then pick a subset of the remaining transmitters such that
their total number of antennas is the closest to the number of antennas of the
receiver set, namely $gN$. Such grouping creates a two users MIMO interference
channel to which the known DoF region will be applied.

Consider an arbitrary subset of receivers $G_{R_1} \subseteq \cK$ with
cardinality $g$. Let $G_{T_1}=G_{R_1}$. The set $G_{T_1}$ contains indices of
transmitters whose signals are useful for the receivers in $G_{R_1}$. We
define another subset of transmitters, $G_{T_2}\subseteq \cK \setminus
G_{T_1}$, such that
\begin{enumerate}
\item The cardinality of $G_{T_2}$ is $\min\{K-g,\lfloor \frac{gN}{M}
\rfloor\}$.
\item Set $G_{T_2}$ maximizes $\sum_{k\in G_{T_2}}{d_k}$.
\end{enumerate}
The corresponding receivers of $G_{T_2}$ are shown by set $G_{R_2}$. We then
remove all the remaining users with indices in $\cK\setminus \{G_{T_1}\cup
G_{T_2}\}$.

\begin{theorem}  \label{th.outer}
For the aforementioned $G_{T_1}$, $G_{T_2}$, and $g$, the following equations
define a DoF region outer bound
for the $(K,[M],[N])$ interference channel:
\begin{align}
\sum_{k\in G_{T_1}}{d_k}&\leq g\min{(M,N)} \label{outer1} \\ \sum_{\hk\in
G_{T_2}}{d_{\hk}}&\leq \min\{K-g,\lfloor \frac{gN}{M} \rfloor\}
\min{(M,N)} \label{outer2}\\ \sum_{k\in G_{T_1}}{d_k} + \sum_{\hk\in
G_{T_2}}{d_{\hk}} &\leq gN. \label{outer3}
\end{align}
\end{theorem}

\emph{Proof:} In \cite{jafa07}, it is proved that the DoF region for a
$2$-user MIMO Gaussian interference channel with $M_1$, $M_2$ antennas at
transmitters and $N_1$, $N_2$ antennas at the corresponding receivers is
\begin{IEEEeqnarray}{Cl}
d_1\leq \min{(M_1,N_1)},\quad d_2\leq \min{(M_2,N_2)}\nonumber \\ d_1+d_2\leq
\min\{M_1+M_2, N_1+N_2,\max{(M_1,N_2)}, \max{(M_2,N_1)}\}
\end{IEEEeqnarray}
Using this result when $G_{T_1}$, $G_{R_1}$ are viewed as the first user and
$G_{T_2}$, $G_{R_2}$ as the second user, we arrive at
\eqref{outer1}--\eqref{outer3}.\hfill \QED

\remark By considering all $1\leq g\leq K$ and for each $g$ all possible
$G_{T_1}\subseteq \cK$ with cardinality $g$, the outer bound can be optimized.

As a special case, if we set all $d_k$ equal to $d$, we have
\begin{equation}
gd + \min\{K-g,\lfloor \frac{gN}{M} \rfloor\} d \leq gN
\end{equation}
for all $g\in\cK$. The above inequality can be represented as
\begin{equation}
d \leq {\frac{gN}{\min\{K,\lfloor \frac{g(N+M)}{M} \rfloor\}}}.
\end{equation}
Therefore, the outer bound for the total DoF is obtained as
\begin{equation}
\min_{g\in\cK}{\frac{gNK}{\min\{K,\lfloor \frac{g(N+M)}{M} \rfloor\}}}.
\end{equation}

For $K \geq \frac{M+N}{\text{gcd}(M,N)}$, we are able to choose
$g=\frac{M}{\text{gcd}(M,N)}$ resulting in the same number of antennas at
transmitters in $G_{T_2}$, and at receivers in $G_{R_1}$. Subsequently, the
total DoF is upper bounded by $\frac{MN}{M+N}K$, which is achievable according
to \thref{th.KMN}.

It can be seen that having an identical number of antennas at the receive side
of user 1 and transmit side of user 2 is important for establishing the
optimality of total DoF. In other words, the desired outer bound occurs when
the receivers of group user 1 with $gN$ antennas are able to successfully
decode interferences created by $gN$ antennas. Such requirement can be
satisfied if $K \geq \frac{M+N}{\text{gcd}(M,N)}$.

\remark Zero-forcing always allows us to achieve the total DoF
$\min\{\max{(M,N)},K\min{(M,N)}\}$, which is indeed tight when
$K<\frac{M+N}{\min(M,N)}$, cf.~\cite{ghmk10c}.

\remark \label{outer.KNN} In the case $M=N$, it is optimal to set $g=1$.
Therefore, the DoF region is upper bounded by
\begin{equation}
d_k + \max_{\hk\in\cK, \hk\neq k}{d_{\hk}} \leq N
\end{equation}
for all $k\in\cK$.

To improve outer bounds associated with grouping approach, a new method in
\cite{wasj12} called genie chains is proposed where a receiver is provided
with a subspace of signals (part of transmitted symbols) as a genie. As a
result of this approach, the total DoF $\frac{MN}{M+N}$ is obtained for the
wider range of $\frac{M}{N}\geq\frac{K-2}{K^2-3K+1}$.

In MIMO X network channel, a general outer bound has been obtained in
\cite{caja09}. It is shown that the sum of all the DoFs of the messages
associated with transmitter $k$ and receiver $j$ is upper bounded by
$\max{(M_k,N_j)}$. Despite the assurance that the total DoF outer bound is
achieved for the single antenna X network, the characterization for the case
of MIMO seems to be challenging.

\section{Conclusions and Future Works}\label{future}

We developed a new real interference alignment scheme for multiple-antenna
interference networks that employed joint receiver antenna processing. The
scheme utilized a result on simultaneous Diophantine approximation and aligned
all interferences at each receive antenna. We were able to derive several new
DoF region results, as summarized in the theorems.

It is desirable to extend the result of the paper to a multiple-antenna
interference network with $K$ transmitters and $J$ receivers where each
transmitter sends an arbitrary number of messages, and each receiver may be
interested in an arbitrary subset of the transmitted messages. The asymptotic
alignment schemes have been successfully used to achieve the optimal DoF for
both SISO and MIMO wireless networks for time-varying channels. It is
interesting to translate these result to the constant channels under real
interference alignment framework and find the connection between real and
vector interference alignment. It is also possible that one can improve the
existing outer bounds so that the optimality of the achieved DoF regions are
generally proved.

\medskip \noindent \emph{Acknowledgment}: The authors thank V. Beresnevich for
comments on the convergence problem of Diophantine approximation on manifolds
and directing us to reference \cite{klmw10}.

\appendix

\subsection{Nondegenerate manifolds} \label{ap.nondegen}

One important notion in studying Diophantine approximation on manifolds is the
so called nondegeneracy, which we briefly review the useful definitions and
facts; see \cite{klma98,bere02} for more discussion.

A smooth map $\bf$ from $U\subset \setR^d$ to $\setR^m$ is called
\emph{$l$-nondegenerate} at $\bx\in U$ if partial derivatives of $\bf$ at
$\bx$ up to order $l$ span $\setR^m$. The mapping $\bf$ is called
\emph{non-degenerate} if for almost every $\bx\in U$ it is $l$-nondegenerate
for some $l$. The non-degeneracy of a manifold guarantees that the manifold
can not be approximated by a hyperplane ``too well''; see
\cite[Lemma~1]{bere02}.

A set of functions are \emph{linearly independent over $\setR$} if none of the
functions can be represented by a linear combination of the other functions
with real coefficients. If the functions $f_1,\ldots,f_n$ are analytic, and
$1,f_1,\ldots,f_n$ are linearly independent over $\setR$ in a domain $U$, all
points of $\cM=\bf(U)$ are nondegenerate.

\subsection{Proof of \leref{le.manifold}} \label{ap.manifold}

In the following, we will need the concept of \emph{strongly extremal},
\emph{very well multiplicative approximable} (VWMA), and \emph{very well
approximable} (VWA). For definitions of these concepts, we refer the reader to
\cite[Sec.~1]{klmw10}.

Based on \cite[Thoerem~1.2]{klmw10}, the pushforward of Lebesgue measure on
$U_1\times\ldots\times U_n$ by $\bA$ is strongly extremal. That is, for almost
all $(\bh_1, \ldots,\bh_n)$, $\bA(\bh_1, \ldots,\bh_n)$ is not VWMA, which in
turn implies that $\bA$ is not VWA. The fact $A$ is not VWA means that there
are at most finitely many $\bq\in \setZ^m$ with
\begin{equation}\label{eq.bound}
  \|\bA\bq-\bp\|_\infty <\|\bq\|_\infty^{-m/n-\delta} \text{ for some } \bp \in \setZ^n
\end{equation}
We require $\bA$ to have at least one row whose elements are rationally
independent, so that for any non-zero $\bq$, $\|\bA\bq\|_\infty >0$. For such
$\bA$ and for all the $\bq$ such that \eqref{eq.bound} does not hold, knowing
that there are at most finitely many of such $\bq$, it is possible to choose
$Q$ large enough such that $\|\bA\bq\|_\infty > Q^{-m/n-\delta}$. As a result,
for large enough $Q$, for all $\bq \in \setZ_Q^m$, we have $\|\bA\bq\|_\infty
> Q^{-m/n-\delta}$. Since the 2-norm is at least as large as the infinity
norm, the desired result is obtained. \hfill \QED

\subsection{Proof of \leref{le.dof}} \label{ap.dof}

The proof is similar to that in \cite{mgmk09}. The difference here is that it
does not resort to the Fano's inequality. Without loss of generality, we fix
the average power per symbol to be $P_0$ and set the per-element noise
variance to 1. Let $w\bydef m/n$, which measures the ratio of the width and
height of matrix $\bA$. Fix $0<\epsilon<1$ and
$0<\delta<\frac{\epsilon(1+w)}{1-\epsilon}$. For large enough $P_0$, we select
$\bx\in \lambda\setZ^m_{Q}$, where
\begin{equation}
  Q={P_0}^{\frac{1-\epsilon}{2(1+w)}}, \quad \lambda=\frac{{P_0}^{1/2}}{Q}={P_0}^{
  \frac{w+\epsilon}{2(1+w)}}
\end{equation}
From \leref{le.manifold}, we know that for almost all $\bA$, and for all $\bx,
\bx'\in \setZ_Q^m$, such that $\bx\ne \bx'$, we have
\begin{equation}\label{eq.pwd}
\|\bA(\bx-\bx')\|_2>d_\text{min}\bydef \lambda (2Q)^{-m/n-\delta} =
2^{-m/n-\delta}{P_0}^{\frac{\epsilon}2 - \frac{\delta(1-\epsilon)}{2(1+w)}}.
\end{equation}
By the choice of $\delta$, the pairwise distance in \eqref{eq.pwd} grows with
$P$ as $P\to\infty$. The pairwise error probability is therefore upper bounded
by
\begin{equation}
  \int_{d_\text{min}/2}^\infty \frac 1{\sqrt{2\pi}} \exp(-t^2/2) dt
  \le \exp(-d_\text{min}^2/8)
\end{equation}
where the Chernoff bound for the Gaussian Q-function has been applied.
Employing the union bound, we can upper bound the average probability of error
as
\begin{align}
  P_e  &< (2Q+1)^m \exp(-d_\text{min}^2/8)\\
  & < (3Q)^m \exp(-d_\text{min}^2/8) \\
  & =3^m \exp\left [ m\frac{1-\epsilon}{2(1+w)} \log{P_0} - \frac 18 \cdot 2^{-\frac{2m}n
  -2\delta}{P_0}^{\epsilon-\frac{\delta(1-\epsilon)}{2(1+w)}} \right].
\end{align}
By the choice of $\delta$, the exponent of $P_0$, namely
$\epsilon-\frac{\delta(1-\epsilon)}{2(1+w)}$ is positive. Also for large
$P_0$, the polynomial term dominates the $\log(P_0)$ term in the exponent. As
a result, the upper bound goes to zero as $P_0\to \infty$. The achieved DoF
per symbol is
\begin{equation}\label{DoFPS}
 \lim_{P_0\to\infty} \frac{\log(2Q+1)}{0.5\log(P_0)} = \frac{1-\epsilon}{1+w}.
\end{equation}
Since $\epsilon$ can be made arbitrarily small, the per-symbol DoF of
$1/(1+w)=n/(m+n)$ can be achieved. The total achieved DoF is $mn/(m+n)$.
\hfill \QED

\linespread{1.3}\normalsize
\bibliography{refs}
\bibliographystyle{IEEE-unsorted}

\end{document}